\documentclass[10pt]{elsarticlehome}
\usepackage{epsfig}

\begin{document}

\title{Emergent Horava gravity in graphene}

\author[AU,ITP]{G.E.~Volovik}

\author[ITEP]{M.A.~Zubkov
%\footnote{Corresponding author, e-mail: zubkov@itep.ru}
 }

\address[AU]{Low Temperature Laboratory, School of Science and
Technology, Aalto University,  P.O. Box 15100, FI-00076 AALTO, Finland}

\address[ITP]{L. D. Landau Institute for Theoretical Physics,
Kosygina 2, 119334 Moscow, Russia}

\address[ITEP]{ITEP, B.Cheremushkinskaya 25, Moscow, 117259, Russia
}

\begin{abstract}
First of all, we reconsider the tight - binding model of monolayer graphene, in which the
variations of the hopping parameters are allowed. We demonstrate that the emergent $2D$ Weitzenbock geometry as well as the emergent $U(1)$ gauge field appear. The emergent gauge field is equal to the linear combination of the components of the zweibein. Therefore, we actually deal with the gauge fixed version of the emergent $2+1$ D teleparallel gravity.  In particular, we work out the case, when the variations of the hopping parameters are due to the elastic deformations, and relate the elastic deformations with the emergent zweibein. Next, we investigate the tight - binding model with the varying intralayer hopping parameters for the multilayer graphene with the $ABC$ stacking. In this case the emergent $2D$ Weitzenbock geometry and the emergent $U(1)$ gauge field appear as well, the emergent low energy effective field theory has the anisotropic scaling.
\end{abstract}

%%%%%%%%%%%%%%%%%%%%%%%%%%%%%%%%%%%%%%%%%%%%%%%%%%%%%%%%%%%%%%%%%%%%%%%%

%%%%%%%%%%%%%%%%%%%%%%%%%%%%%%%%%%%%%%%%%%%%%%%%%%%%%%%%%%%%%%%%%%%%%%%

\maketitle

%\documentclass[a4paper,11pt]{article}
%\pdfoutput=1 % if your are submitting a pdflatex (i.e. if you have
             % images in pdf, png or jpg format)

%\usepackage{jheppub} % for details on the use of the package, please
                     % see the JHEP-author-manual

%\usepackage[T1]{fontenc} % if needed

%\usepackage{slashed}
%\usepackage[dvips]{color}
%\usepackage[dvips]{epsfig}
%\usepackage{latexsym}
%\usepackage{bm}
%\usepackage{upgreek}
%\usepackage{mathrsfs}
%\usepackage{times}
%\usepackage{amsthm}
%\usepackage{amssymb}
%\usepackage{epsfig}
%\usepackage{graphicx}
%\usepackage{amsmath}

\newcommand{\barray}{\begin{eqnarray}}
\newcommand{\earray}{\end{eqnarray}}
\newcommand{\nn}{\nonumber \\}
\newcommand{\nl}{& \nonumber \\ &}
\newcommand{\bnl}{\right .  \nonumber \\  \left .}
\newcommand{\dbnl}{\right .\right . & \nonumber \\ & \left .\left .}

%Begin-end
\newcommand{\beq}{\begin{equation}}
\newcommand{\eeq}{\end{equation}}
\newcommand{\ba}{\begin{array}}
\newcommand{\ea}{\end{array}}
\newcommand{\bea}{\begin{eqnarray}}
\newcommand{\eea}{\end{eqnarray} }
\newcommand{\be}{\begin{eqnarray}}
\newcommand{\ee}{\end{eqnarray} }
\newcommand{\bal}{\begin{align}}
\newcommand{\eal}{\end{align}}
\newcommand{\bi}{\begin{itemize}}
\newcommand{\ei}{\end{itemize}}
\newcommand{\ben}{\begin{enumerate}}
\newcommand{\een}{\end{enumerate}}
\newcommand{\bc}{\begin{center}}
\newcommand{\ec}{\end{center}}
\newcommand{\bt}{\begin{table}}
\newcommand{\et}{\end{table}}
\newcommand{\btb}{\begin{tabular}}
\newcommand{\etb}{\end{tabular}}
\newcommand{\bvec}{\left ( \ba{c}}
\newcommand{\evec}{\ea \right )}

\newcommand\e{{e}}
\newcommand\eurA{\eur{A}}
\newcommand\scrA{\mathscr{A}}

\newcommand\eurB{\eur{B}}
\newcommand\scrB{\mathscr{B}}

\newcommand\eurV{\eur{V}}
\newcommand\scrV{\mathscr{V}}
\newcommand\scrW{\mathscr{W}}

\newcommand\eurD{\eur{D}}
\newcommand\eurJ{\eur{J}}
\newcommand\eurL{\eur{L}}
\newcommand\eurW{\eur{W}}

\newcommand\eubD{\eub{D}}
\newcommand\eubJ{\eub{J}}
\newcommand\eubL{\eub{L}}
\newcommand\eubW{\eub{W}}

\newcommand\bmupalpha{\bm\upalpha}
\newcommand\bmupbeta{\bm\upbeta}
\newcommand\bmuppsi{\bm\uppsi}
\newcommand\bmupphi{\bm\upphi}
\newcommand\bmuprho{\bm\uprho}
\newcommand\bmupxi{\bm\upxi}

\newcommand\calJ{\mathcal{J}}
\newcommand\calL{\mathcal{L}}

\newcommand{\notyet}[1]{{}}

\newcommand{\sgn}{\mathop{\rm sgn}}
\newcommand{\tr}{\mathop{\rm Tr}}
\newcommand{\rk}{\mathop{\rm rk}}
\newcommand{\rank}{\mathop{\rm rank}}
\newcommand{\corank}{\mathop{\rm corank}}
\newcommand{\range}{\mathop{\rm Range\,}}
\newcommand{\supp}{\mathop{\rm supp}}
\newcommand{\p}{\partial}
\renewcommand{\P}{\grave{\partial}}
\newcommand{\yDelta}{\grave{\Delta}}
\newcommand{\yD}{\grave{D}}
\newcommand{\yeurD}{\grave{\eur{D}}}
\newcommand{\yeubD}{\grave{\eub{D}}}
\newcommand{\at}[1]{\vert\sb{\sb{#1}}}
\newcommand{\At}[1]{\biggr\vert\sb{\sb{#1}}}
\newcommand{\vect}[1]{{\bold #1}}
\def\R{\mathbb{R}}
\newcommand{\C}{\mathbb{C}}
\def\hvar{{\hbar}}
\newcommand{\N}{\mathbb{N}}\newcommand{\Z}{\mathbb{Z}}
\newcommand{\Abs}[1]{\left\vert#1\right\vert}
\newcommand{\abs}[1]{\vert #1 \vert}
\newcommand{\Norm}[1]{\left\Vert #1 \right\Vert}
\newcommand{\norm}[1]{\Vert #1 \Vert}
\newcommand{\Const}{{C{\hskip -1.5pt}onst}\,}
\newcommand{\sothat}{{\rm ;}\ }
\newcommand{\Range}{\mathop{\rm Range}}
\newcommand{\ftc}[1]{$\blacktriangleright\!\!\blacktriangleright$\footnote{AC: #1}}

% Italic ``theorems''
%\theoremstyle{plain}
\newtheorem{lemma}{Lemma}[section]
\newtheorem{hypothesis}[lemma]{Hypothesis}
\newtheorem{corollary}[lemma]{Corollary}
\newtheorem{proposition}[lemma]{Proposition}
\newtheorem{claim}[lemma]{Claim}

\newtheorem{theorem}[lemma]{Theorem}

% Roman ``theorems''
%\theoremstyle{definition}
\newtheorem{definition}[lemma]{Definition}
\newtheorem{assumption}[lemma]{Assumption}

% Humble things: remarks and examples.
%\theoremstyle{remark}
\newtheorem{remark}[lemma]{Remark}
\newtheorem{example}[lemma]{Example}
\newtheorem{problem}[lemma]{Problem}
\newtheorem{exercise}[lemma]{Exercise}

\newcommand{\const}{\mathop{\rm const}}

\renewcommand{\theequation}{\thesection.\arabic{equation}}

\makeatletter\@addtoreset{equation}{section}
%\makeatletter\@addtoreset{theorem}{section}
%\makeatletter\@addtoreset{proposition}{section}
%\makeatletter\@addtoreset{lemma}{section}
%\makeatletter\@addtoreset{corollary}{section}
%\makeatletter\@addtoreset{remark}{section}
%\makeatletter\@addtoreset{assumption}{section}
%\makeatletter\@addtoreset{definition}{section}
\makeatother

\def\Tau{\mathcal{T}}

\def\os{{o}}
\def\ol{{O}}
\def\dist{\mathop{\rm dist}\nolimits}
\def\spec{\sigma}
\def\mod{\mathop{\rm mod}\nolimits}
\renewcommand{\Re}{\mathop{\rm{R\hskip -1pt e}}\nolimits}
\renewcommand{\Im}{\mathop{\rm{I\hskip -1pt m}}\nolimits}

%\input{t1ptm.fd}
%\input{ulasy.fd}
%\input{ueus.fd}
%\input{ueuf.fd}

%\input{ursfs.fd}

%\title{\boldmath Emergent Horava gravity in graphene}

%% %simple case: 2 authors, same institution
% \author{G.E. Volovik$^{a,b}$,}
%% \author{and A. Nother Author}
% \affiliation{$^{a}${ Olli Lounasmaa Laboratory, School of Science and Technology,
%Aalto University, Finland} \\
%$^{b}${ L.D. Landau Institute for Theoretical Physics, Moscow, Russia}}

% more complex case: 4 authors, 3 institutions, 2 footnotes
%\author{M.A.Zubkov$^{c}$}

% The "\note" macro will give a warning: "Ignoring empty anchor..."
% you can safely ignore it.

%\affiliation{$^{c}$Institute for Theoretical and Experimental Physics, Moscow, Russia}

% e-mail addresses: one for each author, in the same order as the authors
%\emailAdd{zubkov@itep.ru}

%\begin{document}
%\maketitle
%\flushbottom

\def\w{\omega}
\def\o{\omega}

\newcommand{\br}{{\bf r}}
\newcommand{\bu}{{\bf \delta}}
\newcommand{\bk}{{\bf k}}
\newcommand{\bq}{{\bf q}}
\def\({\left(}
\def\){\right)}
\def\[{\left[}
\def\]{\right]}

\section{Introduction}

In the fermionic systems in the vicinity of the Fermi points the emergent
Lorentz symmetry appears
\cite{Horava2005,NielsenNinomiya1981,Froggatt1991,Volovik2003,Volovik2011}.
It was suggested in \cite{Horava2005} that this is related to the
Atiyah-Bott-Shapiro construction applied to momentum space topology.
According to the conjecture of \cite{Horava2005} the first (linear) term in
expansion of the effective fermionic action near the Fermi point in $3+1$ D
is expressed  in
terms of  Pauli matrices:
\begin{equation}
{\cal S} = \int d^{4} x |{\rm det}\,e| \, e^{\mu}_a \, \bar{\psi}
\sigma^a(p_\mu- {\cal A}_\mu)\psi +  \ldots \label{Hamiltonian}
\end{equation}

This means  that near the Fermi point, i.e. at low energy, the fermionic
excitations behave as Weyl particles. The parameters of
expansion become the dynamical bosonic fields, where the collective mode
${\cal A}_\mu$ may contain the shift of the Fermi point resulted in the
effective gauge field $A_{\mu}$ and the effective spin connection
$\frac{i}{8} C^{ab}_{\mu} [\sigma^a,\sigma^b]$. Under certain circumstances
the modes $e^{\mu}_a$ behave as the vielbein describing the gravitational
degrees of freedom
with the effective metric field
 $g_{\mu\nu}=e^{\mu}_ae^{\nu}_b \eta^{ab}$.
It is worth mentioning that depending on the particular properties of the
system under consideration the fields $e^a_i, A_{\mu}, C^{ab}_{\mu}$ may
depend on each other. In most of the cases the  main symmetry of the
gravitational theory (invariance under the diffeomorphisms) does not arise.
If the fields $e^a_i, A_{\mu}, C^{ab}_{\mu}$ are not independent, the
emergent low energy action of the form of Eq. (\ref{Hamiltonian}) contains
the terms that are not invariant under the diffeomorphisms. This gives the
principal difficulty. However, this difficulty is avoided if Eq.
(\ref{Hamiltonian}) is considered as a gauge fixed version of the action for
the theory invariant under the diffeomorphisms.
If $C^{ab}_{\mu}$ is absent or is expressed through the emergent vielbein $e^a_k$, the
given construction results in the teleparallel gravity, i.e. the theory of
the varying Weitzenbock geometry\footnote{The Riemann - Cartan space is
defined by the translational connection (the zweibein) and the Lorentz group
connection. There are two important particular cases. space is called
Riemannian if the translational curvature (torsion) vanishes. If the Lorentz
group curvature vanishes, it is called Weitzenbock space.}. If
$C^{ab}_{\mu}$ varies independently of the vielbein, this may give the
scenario of how topology leads to the emergence of Riemann - Cartan gravity.

 Condensed matter example of the emergent gravity
is provided by Weyl syperfluid $^3$He-A \cite{Volovik2011}. Strictly
speaking, the construction of \cite{Horava2005} works for the odd dimension
of space. Here we demonstrate, that the similar construction works
also in monolayer graphene that is the $2+1$ system (see also \cite{Vozmediano2010}).
In the non-topological systems, the relativistic
tetrad gravity may also arise, see e.g.
\cite{Akama1978,Volovik1986,Wetterich2004,Diakonov2011}, but it is not
universal, since is not accompanied by the emergence of spinor and vectors
fields.

The Horava construction \cite{Horava2005} refers to nodes with the
elementary topological charge
$N=\pm 1$ in the so-called stable regime. In case of the degenerate nodes
and for the other non-typical nodes the symmetry of the  underlying
microscopic system becomes instrumental. There are many examples where
topology in combination with the symmetry consideration gives rise to exotic
spectrum,
with nonlinear touching of energy branches at Fermi point (such as
quadratic, cubic, quartic, etc.  touching, i.e. $E^2=p^{2J}$ with $J=2,3,4
\ldots$
\cite{HeikkilaVolovik2010,HeikkilaVolovik2011}; and the
mixed linear and quadratic touching $E^2=p_z^{2}+ p_\perp^4$
\cite{Volovik2003}).

The quantum field theory (QFT) of fermions and bosons, which emerges in the
vicinity of an exotic Fermi point, is certainly non-relativistic. But this
is the subject of the QFT with the anisotropic scaling
${\bf r} \rightarrow b{\bf r}$, $t \rightarrow b^z t$,
which recently got attention due to the construction of the so-called
Ho\v{r}ava-Lifshitz gravity
\cite{HoravaPRL2009,HoravaPRD2009,Horava2008,Horava2010}.
The QED with anisotropic scaling emerging in such systems has been discussed
in Refs.
\cite{KatsnelsonVolovik2012,KatsnelsonVolovikZubkov2013a,KatsnelsonVolovikZubkov2013b}.
However, when trying find the emergent Ho\v{r}ava-Lifshitz gravity, one
encounters the problem.
When the construction of \cite{Horava2005} is generalized to the non-linear
touching, the
natural  expansion
$e^a_{ijk...l} p^I p^j p^k ... p^l \gamma^a$  describing the gravitational
degrees of freedom contains  parameters $e^a_{ijk...l}$ instead of the
vielbein $e^{\mu}_a$. This expansion does not produce the metric field, and
thus cannot serve as a possible microscopic source of Ho\v{r}ava-Lifshitz
gravity.

In addition to the system with the isotropic scaling (monolayer graphene) we consider the particular $2+1$ D topological system (multilayer
graphene with ABC - stacking) with the emergent anisotropic scaling for
the non-relativistic Dirac quasiparticles.  We consider the situation, when
the parameters of the tight - binding model vary. Due to the symmetry
specific for the given system the effective hamiltonian contains the
emergent zweibein, the $2D$ spin connection, and the $U(1)$ gauge field.
Thus, the variations of the parameters of the hamiltonian lead to the
emergent Ho\v{r}ava-Lifshitz gravity, and produce the
components of vielbein, spin connection, and $U(1)$ gauge field.

In multilayer graphene with $ABC$ stacking the Fermi
point emerges with the quasiparticle spectrum $E^2=p^{2J}$. This is a stack
of  $J$ identical membranes, which  interact in a manner discussed in Refs.
\cite{HeikkilaVolovik2010,HeikkilaVolovik2011}. If the interaction between
the branes  is ignored, the spectrum in each membrane contains topologically
protected Fermi point with $N=1$. The quasiparticles in each brane are
relativistic in the low-energy limit. This is similar to the Horava
construction in the systems with
 odd dimension of space and gives the linear expansion near the Fermi point
$e^{\mu}_a \gamma^a(p_\mu-
{\cal A}_\mu)+  \ldots$.

When the interaction between the membranes is switched on, the original
nodes at $p=0$ with $N=1$ on  independent $J$ branches are combined to a
single node with the topological charge $N=J$  and with the spectrum
$E^2\sim(g^{ik}p_i p_k)^{J}$. The rest $J-1$ branches become gapped. In
this mechanism the gapless branch of exotic Dirac fermions inherits the
original vielbein structure of the individual branes
$g^{ik}=e^{i}_ae^{k}_b \delta^{ab}$.  This mechanism may serve
as a microscopic source of the Ho\v{r}ava-Lifshitz gravity satisfying the
anisotropic scaling  with $z=J$.

The reason, why the nonlinear spectrum of exotic fermions remembers the
original tetrad fields of individual branes, is pure topological.  The role
of topology can be demonstrated in the limit of large $J$. For large $J$,
the gapless branch is concentrated on the outer branes in the stack. In the
limit $J\rightarrow \infty$, the stack of branes transforms to the 3+1  bulk
topological material of the class of  the topological semimetals with the
nodal lines \cite{HeikkilaVolovik2011,SchnyderRyu2010}). The branch with
gapless spectrum transforms to the branch of edge states emerging on the
surface of this material. In topological materials the edge states are not
fully autonomous: they reflect the topological properties of bulk state and
their degrees of freedom are restricted (see e.g.
\cite{HasanKane2010,Xiao-LiangQi2011,Burkov2011,XiangangWan2011,Kaplan2012}).
The Ho\v{r}ava-Lifshitz gravity (see also \cite{Horava2}) is the outcome of the momentum-space
topology in the mesoscopic  finite $J$ regime. This is the particular
example of how the
low energy degrees of freedom are restricted due to symmetry and topology of
the underlying microscopic physics.

\section{Varying parameters of the tight - binding model for monolayer
graphene}

\label{sectmono1}

%In \cite{Vozmediano,Vozmediano1} it was considered how the elastic deformations affect the tight - binding model of monolayer graphene.
Here we consider the variations of the hopping parameters of general type, that are not necessarily related to the elastic deformations.
The low energy effective model of graphene may be derived
\cite{Semenoff:1984dq,Wallace,CastroNeto:2009zz} starting from the simple
non -
relativistic Hamiltonian that describes the interactions of electrons that
belong to neighbor Carbon atoms. The carbon atoms of graphene form a
honeycomb
lattice with two sublattices A and B (of the triangular form). We
denote the lattice spacing by $a$. Let us introduce vectors that connect a
vertex of the sublattice A to its neighbors (that belong to the sublattice
B): ${\bf l}_1= (-a,0)$, ${\bf l}_2 =  (a/2,a\sqrt{3}/2)$, ${\bf l}_3=
(a/2,-a\sqrt{3}/2)$.

Normally the Hamiltonian depends on one hopping parameter and describes the
jumps between the adjacent sites of the honeycomb lattice. Operator
$\psi^\dag$ is introduced that creates electrons and annihilates holes at
the points of the lattice.
Let us suppose that the hopping parameter varies, so that its value
depends on the particular link connecting two adjacent points of the
honeycomb lattice. This may be caused by elastic deformations as in
\cite{Vozmediano}, but the other independent variations of the hopping
parameters may also appear (for example in the artificial molecular graphene
\cite{Gomes2012}). We have three values of $t_a, a = 1,2,3$ at each
point.
The Hamiltonian has the form
\begin{equation}
H=-\sum_{\alpha\in A}\sum_{j=1}^3 t_j(\br_\alpha)
    \Bigl(\psi^\dag (\br_\alpha) \psi(\br_\alpha + {\bf l}_j)
        + \psi^\dag (\br_\alpha +{\bf
l}_j)\psi(\br_\alpha)\Bigr)\,,\label{H12}
\end{equation}

In Appendix this Hamiltonian is considered in details. We demonstrate that in the case of the small variations of $t_a$ that do not depend on the coordinates there still exist two Fermi points, and near these Fermi points the two spinors $\Psi_{\pm}$ appear. The  hamiltonians ${\bf H}_{\pm}$ corresponding to the two valleys are related by the charge conjugation:  ${\bf H}_{+}({\bf A}) = \sigma^2 {\bf H}_-(-{\bf A})\sigma^2$. (Here $\bf A$ is the emergent  $U(1)$ gauge field.) Because of this relation we may consider the effective low energy theory for the quasiparticles living near to the single valley. We choose for the definitness the valley $K_-$ with the  low energy effective  hamiltonian
\begin{eqnarray}
H &=&  -\frac{i}{2} \int  e\, d^2 x\Bigl( \bar{\Psi}_{-}({\bf x})  {\bf e}_a^k \sigma^a D_k
\Psi_{-}({\bf x}) - [D^{\dag}_k\bar{\Psi}_{-}({\bf x})]  {\bf e}_a^k \sigma^a
\Psi_{-}({\bf x})\Bigr)\nonumber\\  &=&  \int d^2 x \bar{\Psi}_{-}({\bf x}) {\cal H}
\Psi_{-}({\bf x})\label{Hamiltonian20}
\end{eqnarray}
Here $\bar{\Psi} = - i \Psi^{\dag}\sigma^3$, and
\begin{eqnarray}
{\cal H}= i \sigma^3 {\bf H}_-  =   - i { e}\, {\bf e}_a^k \sigma^a
\circ [\partial_k + i {\bf A}_k] ,
\label{Hpm}
\end{eqnarray}
where $e \, {\bf e}^k_a \circ i \partial_k = \frac{i}{2} \Bigl( e\, {\bf e}^k_a  \overrightarrow{\partial_k} - \overleftarrow{\partial_k} e\, {\bf e}^k_a \Bigr)$.
We define $e$ in such a way that ${\rm det}\, {\bf e} = 1$. The emergent
zweibein $\bf e$ and the emergent $U(1)$ field $\bf A$ are expressed through the original parameters $t_a$ given by
\begin{equation}
t_a = t (1 - \Delta_a), \quad |\Delta_a| \ll 1
\end{equation}
The explicit dependence is as follows:
\begin{eqnarray}
{\bf e}_a^i &=& {\bf f}_a^i/e, \quad e = [{\rm det}\, {\bf f}]^{1/2} = v_F(1- \frac{1}{3}(\Delta_2 + \Delta_3 + \Delta_1))\nonumber\\
{\bf f}^i_a &=& v_F\left(\delta^i_a -  \left[\begin{array}{cc} \Delta_{1} &  \frac{(\Delta_2 -  \Delta_3)}{\sqrt{3}} \\
                                                          \frac{(\Delta_2 -  \Delta_3)}{\sqrt{3}} &   \frac{2}{3}(-\frac{1}{2}\Delta_1+\Delta_2+\Delta_3)    \end{array}\right]\right)\nonumber\\
{\bf A}_1 & = &  \frac{1}{2v_Fa} ({\bf e}^1_2 + {\bf e}^2_1), \quad {\bf A}_2 = \frac{1}{ 2v_Fa} ({\bf e}^1_1 - {\bf e}^2_2)\label{All}
\end{eqnarray}
Here we define the unperturbed Fermi velocity as $v_F = \frac{3 t}{2a}$.
The basis is chosen in such a way that the first (X) axis is directed along $-{\bf l}_1/a =  (1,0)$ while the second (Y) axis is directed along $-\frac{{\bf m}_2}{a \sqrt{3}} = (0,1)$. The expression for the emergent $U(1)$ gauge field obtained here was already considered earlier (see, for example, \cite{sato}).

In tensorial form the fields $\bf A$, $\bf f$ can be written as:
\begin{eqnarray}
{\bf A}^b &=&  - \frac{2}{3a^2} \epsilon^{ba} \sum_j  \Delta_{j} {\bf l}_j^a\nonumber\\
{\bf f}^k_{a}& = &    v_F \Bigl( \delta^k_a  - \frac{2}{3a^2} \sum_j  \Delta_{j} \Bigl[{\bf l}_j^a {\bf l}_j^k    - \frac{a}{2} {\bf l}^d_j \, K^{dak}\Bigr] \Bigr)
\end{eqnarray}
Tensor $K$ was introduced in \cite{Vozmediano}. It reflects the structure of the honeycomb lattice and is given by $K^{ijk}=-\frac{4}{3a^3} \sum_b {\bf l}_b^i {\bf l}^j_b {\bf l}_b^k$. Its only nonzero components are: 
$K^{111}=-K^{122}=-K^{221}=-K^{212}=1$.   

The next step would be to consider the variations of $t_a$ dependent on the coordinates. In this situation the additional contributions to the effective Hamiltonian appear that are linear in ${\bf f}({\bf r})$. In these additional terms the derivatives act on the field $\bf f$ instead of the spinors. It is shown in sect. \ref{Appendix5} of Appendix, that these additional terms have the form of the additional  $U(1)$ gauge field $\tilde{\bf A}_a \approx \frac{1}{2 v_F}\nabla_k {\bf f}^k_b\epsilon_{ba}$.
Instead of Eq. (\ref{Hpm}) we have
\begin{eqnarray}
{\cal H}= - i{ e}({\bf x})\, {\bf e}_a^k({\bf x}) \sigma^a \circ
[\partial_k + i {\bf A}_k({\bf x}) + i \tilde{\bf A}_k({\bf x})]
\label{Hpm2}
\end{eqnarray}
We should compare the values of the additional field $\tilde{\bf A}$ with the emergent gauge field $\bf A$ of Eq. (\ref{All}).

The conditions that allow to use the field - theoretical description is that all fields vary slowly, i.e. their variations on the distances of the order of the lattice spacing are small. That is why we assume, in particular, that quantities $\Delta_a$  vary slowly at the distances of the order of the spacing $a$, that is $a|\partial_k \Delta_b| \ll |\Delta_b| \ll 1$.
Varying field ${\bf e}$ generates the new energy scale ${\cal E}_{em}$ that corresponds to the emergent gauge field. It is given by ${\cal E}_{em} \sim \frac{\Delta_b}{a}$. At the same time the scale of the field $\tilde{\bf A}$  is $\tilde{\bf A} \sim  a \nabla {\bf A}$. This means that we cannot keep this additional field together with $\bf A$.
That's why even in case of the variations of $t_a$ depending on the position in coordinate space we are left with the Hamiltonian of Eq. (\ref{Hpm}) with $\bf e$ and $\bf A$ given by Eq. (\ref{All}). (Now it is assumed that $\Delta_a({\bf x})$ depends on the position in coordinate space $\bf x$.) For the same reason we may substitute the operation ${\bf f}^k_a \circ i \partial_k$ by the usual product ${\bf f}^k_a  i \partial_k$: the difference between the two products is proportional to the derivative of $\bf f$ and could be neglected.
However, in some of the expressions we shall keep the product $\circ$ in order to keep the Hamiltonian manifestly hermitian.

The expression for the field $\bf A$ can also be rewritten as:
\begin{eqnarray}
{\bf A}^i & \approx &  -\frac{1}{2v_Fa}\epsilon^{ik} K^{kjb} {\bf e}^j_b \label{All_inv}
\end{eqnarray}

The important fact about
this dependence is that the emergent field $\bf A$ is expressed through
the zweibein $\bf e$.  So, we have three
independent parameters at each point that are in one to one correspondence
with the components of the $2D$ metric or with the components of the
zweibein considered in a certain gauge. (In the next section we shall point out how
the fields $\bf e, A $ are expressed through the elastic deformations in the case, when they are the source of the varying
hopping parameters $t_a$.)

Thus, the variations of $t_a$ give rise to the $2D$ geometry and the
emergent electro - magnetic field. The
varying $2D$ geometry is given by the zweibein  $\bf e$; the $SO(2)$
connection $\bf C$ disappears in the limit, when we are able to use the field - theoretical description.  Therefore, the geometry is of the Weitzenbock type\footnote{It is worth
mentioning that earlier the hypothesis was suggested, that the emergent
geometry in graphene is Riemannian (see, for example, \cite{Vozmediano0} and
references therein).}.  Vector potential of the emergent electromagnetic
field $\bf A$ also depends on $t_a$ and is not
independent of the zweibein. It gives rise to the emergent magnetic field
orthogonal to the graphene plane and the
emergent in - plane electric field.

One can see, that there is no emergent invariance under the $2D$ reparametrizations in the low energy effective field theory with the Hamiltonian of Eqs. (\ref{Hamiltonian20}) and (\ref{Hpm}). If one assumes the transformation ${\bf e}^i_a \rightarrow \frac{\partial \tilde{x}^i}{\partial x^k}{\bf e}^k_a$, then  $\bf A$  is not transformed as a vector. That's why  Eq. (\ref{Hamiltonian2}) should be considered only as the gauge fixed version of the action invariant under the $2D$ reparametrizations. This gauge corresponds to the choice of the preferred reference frame with the orientation related to the structure of the honeycomb lattice.

Next, let us consider the situation, when the external electromagnetic field is present that compensates exactly the emergent $U(1)$ gauge field considered above for the fermion quasiparticles living near to the Fermi - point $K_-$. (For the fermions living near to $K_+$ it makes the emergent gauge field twice larger.) In this situation the effective Hamiltonian of the system near to $K_-$ receives the form (instead of Eq. (\ref{Hpm})):
\begin{eqnarray}
{\cal H}= -i { e}\, {\bf e}_a^k({\bf x}) \sigma^a
\circ \partial_k
\label{HpmnoA}
\end{eqnarray}

So, in the case, when the emergent $U(1)$ gauge field is exactly compensated by the external electromagnetic field for the fermions living near to one of the Fermi - points, we are left with the pure emergent Weitzenbock geometry defined by the emergent zweibein (near to this Fermi - point). Remarkably, in this case Eqs. (\ref{Hamiltonian20}) and (\ref{HpmnoA}) obey the emergent invariance under the $2D$ reparametrizations.

\section{Elastic deformations as a source of emergent gravity in monolayer graphene}
\label{sectelastic}

In this section we describe how elastic deformations affect the emergent geometry and emergent $U(1)$ field in graphene.

This is natural to suppose, that with our infinite number of microscopic degrees of freedom, which is proportional to the number of atoms, we can organize any deformation of the original graphene field theory, which simulates any degree of freedom of macroscopic quantum field theory and gravity. One may ask what class of deformations is needed for simulation of some particular fields, such as magnetic field, color field, curvature, torsion, etc. In this section we consider the particular form of the deformation that is caused by elastic deformations, and consider what kind of fields and which components they can reproduce. At our level of consideration we assume that the elastic deformations result only in the variations of the hopping parameters described in the previous section.

 Elasticity theory of crystals is a special type of gravity, which does not interact with matter. The metric, torsion and curvature, which  describe deformation of materials, do not
describe the effective space in which the fermions are propagating. Dirac fermions are propagating in different space described by different gravity - like fields, but their  space certainly depends on the deformation of crystals. The problem is to find the connection between the two spaces.

%\subsubsection{Elasticity theory of crystals with dislocations}

The graphene sheet is parametrized by variable $x_k, k = 1,2$. The classical elasticity theory has the displacements $u_a(x)$ as degrees of freedom ($a=1,2,3$).  The three - dimensional coordinates $y_a$ of the graphene sheet are given by
\begin{eqnarray}
y_k(x)& = & x_k + u_k(x), \quad k = 1,2\nonumber\\
y_3(x)& = &u_3(x)
\end{eqnarray}

At $u_a=0$ the graphene is flat. We do not assume that the values of the deformation are small, so that the further expressions work, for example, for the description of the cylindrical shape of the graphene sheet.
We do not consider disclinations and dislocations.  The emergent metric of elasticity theory is given by
\begin{equation}
g_{ik} = \delta_{ik} + 2 u_{ik},~~u_{ik} = \frac{1}{2}\Bigl(\partial_i u_k +
\partial_k u_i +  \partial_i u_a \partial_k u_a\Bigr), \quad a = 1,2,3,
\quad i,k = 1,2.
\label{Two_deformations}
\end{equation}

According to \cite{Vozmediano} the elastic deformations of graphene in (\ref{Two_deformations}) change the hopping elements, which determine the effective geometry experienced by fermions.
The simplest  connection between the deformations and the hopping elements $t_n$ ($n=1,2,3$), which is allowed by symmetry, is in terms of the three unit vectors $l^a_n = \{(-1,0); (1/2,\sqrt{3}/2); (1/2,-\sqrt{3}/2)\}$:
\begin{equation}
t_a({\bf r})=t[1 - {\beta} u_{ik}({\bf r})  l^i_a l^k_a] \,.
\label{HoppingElements}
\end{equation}
In \cite{Vozmediano} in this expression instead of $u_{ik}$ the value $(u_{ik} + \frac{a}{2}\partial_m u_{ik} l^m_a)$ (that attempts to model the value of $u_{kl}$ in the middle of the given link) is used.
We do not see any reason to keep the additional term with the derivative of $u_{ik}$ because in the limit, when the field - theoretical description is acceptable, this derivative being multiplied by the lattice spacing $a$ is to be neglected. This means that $1\gg |u_{ik}| \gg a|\partial_m u_{ik}|$.
The dimensionless phenomenological parameter $\beta$ is determined by the microscopic physics. As it was mentioned above, the given consideration works for the displacements $u_a$ that are not necessarily small. However, we imply that $\beta |u_{ij}| \ll 1$. This is the requirement that the derivatives of $u_a$ are small being multiplied by $\beta$.

It follows from Eq. (\ref{f20}) of the Appendix, that the emergent geometry and emergent $U(1)$ gauge field are given by Eq. (\ref{All}) with
\begin{eqnarray}
 {\bf f}^i_a & = & v_F\left(\delta^i_a - \beta \left[\begin{array}{cc} u_{11} & u_{21}  \\
                                                            u_{12} &       u_{22}   \end{array}\right]\right) \label{elasticfields}
\end{eqnarray}

This results in the usual expression for the strain - induced electromagnetic field:
\begin{eqnarray}
{\bf A}_1  & = &- \frac{\beta}{a}\,u_{12}\nonumber\\
  {\bf A}_2  & = & \frac{\beta}{2a}\,(u_{22}-u_{11})\label{AFP2_}
\end{eqnarray}

It is worth mentioning that our result for the emergent $U(1)$ gauge field Eq. (\ref{AFP2_}) coincides with the one obtained previously (see \cite{Vozmediano,Vozmediano0,Vozmediano1,Vozmediano2010,VK2010} and references therein).  As for our values of ${\bf f}^k_a$, they differ from the expression for the anisotropic Fermi velocity calculated in \cite{Vozmediano}. This is caused by the different method of the calculations. We expand the Hamiltonian near to the true Fermi point given by the unperturbed Fermi point plus the emergent $U(1)$ gauge field. At the same time the authors of \cite{Vozmediano} expand the Hamiltonian near to the unperturbed Fermi point. The latter procedure seems to us incorrect.

The zweibein is given by
\begin{eqnarray}
 {\bf e}^i_a & = & \left(\delta^i_a(1+ \frac{\beta}{2} u_{aa}) - \beta \left[\begin{array}{cc} u_{11} & u_{21}  \\
                                                            u_{12} &       u_{22}   \end{array}\right]\right) \label{elasticfieldse}
\end{eqnarray}
It is constructed in such a way that the determinant of the zweibein ${\rm det}\, {\bf e}^{(2\times 2)} = 1$.
The $2+1$ volume element $d^{(3)}V = e({\bf r},t) \, d^2 {\bf r} dt$  corresponds to the function
\begin{eqnarray}
 {e} & = & v_F (1 - \frac{\beta}{2} u_{aa})
\end{eqnarray}
This function is related to the $(^0_0)$ - component of the dreibein as ${\bf e}^0_0 = 1/e$. The determinant of the dreibein is equal to $1/e$.

So, the elastic deformations affect the motion of the fermions via Eq. (\ref{Hpm}), with the emergent fields $\bf e, A$, and with vanishing $\bf C$. They are expressed through  elastic deformations according to Eqs. (\ref{elasticfields}), (\ref{AFP2_}). In the special case, when the emergent $U(1)$ gauge field is exactly compensated by the external electromagnetic field we are left with the pure emergent Weitzenbock geometry.

It is worth mentioning that in the presence of dislocations two additional fields appear that contribute to torsion. The total  number of degrees of freedom becomes larger by two than the number of independent degrees of freedom of Weitzenbock gravity.
{ The additional degrees of freedom may become the source of the independent  $SO(2)$ gauge field $\bf C$. } However, we do not consider here this possibility.

%\end{enumerate}

\section{Multilayer graphene with $ABC$ stacking}

In this section we consider the system of $J$ interacting $2D$ branes. The
example of such a system is given by the graphene multilayer with $ABC$
stacking. The tight - binding model leads to the low energy block - diagonal
Hamiltonian.
For example, for $J=4$ we have
\cite{HeikkilaVolovik2010,HeikkilaVolovik2011}:
\begin{equation}
{\cal H}_4({\bf p})=
\left( \begin{array}{cccc}
v_F{\sigma}\cdot{\bf p} &t_{\bot} \sigma^+&0&0\\
t_{\bot}\sigma^-&v_F{\sigma}\cdot{\bf p} &t_{\bot}\sigma^+&0\\
0&t_{\bot}\sigma^-&v_F{\sigma}\cdot{\bf p}&t_{\bot}\sigma^+\\
0&0&t_{\bot}\sigma^-&v_F{\sigma}\cdot{\bf p}
\end{array} \right)
\,.
\label{4x4}
\end{equation}
 If the mixing parameter  $t_{\bot}$ between the neighboring branes is zero,
the
fermions
living on each individual brane are described by $2\times 2$ Dirac
Hamiltonian
 ${\cal H}_1=v_F{\sigma}\cdot{\bf p}=v_F(\sigma_x p_x + \sigma_y p_y) $,
which has the elementary topological charge $N=1$ (see e.g.
Refs.\cite{Volovik2011,ZhaoWang2012}). In this case this is the topological
invariant protected by symmetry. Similar to Horava
construction the gapless branch in each brane has the relativistic  spectrum
$E^2=v_F^2p^2$.
For $t_{\bot}\neq 0$, the important symmetry of  Eq.(\ref{4x4}) is that
matrices
$\sigma^+=(\sigma_x + i \sigma_y)/2$ and $\sigma^-=(\sigma_x - i \sigma_y)/2$, which
participate in mixing of branes,  are on the opposite sides of
the diagonal. This symmetry results in
the low energy branch, which has the topological charge $N=4$ and the
spectrum
$E^2=v_F^8p^8/t_{\bot}^{6}$ with quartic touching.

 The extension of Eq.(\ref{4x4}) to the other values of $J$ is
straightforward. For general values $J>1$ the topological charge is $N=J$
and correspondingly the spectrum of the low energy branch is
$E^2=v_F^{2J}p^{2J}/t_{\bot}^{2(J-1)}$.

%\section{Varying parameters of the tight - binding model for multilayer graphene}

 Next, we consider the tight - binding model of the multilayer graphene, in
which the intralayer hopping parameters $t_a$ may vary while the interlayer
hopping parameter $t_{\bot}$ is fixed.
%This approximation is justified
%because the connection between the layers is much slighter than the
%connection between the sites of the honeycomb lattice within each layer.
%Therefore, values of $t_a$ are much larger than $t_{\bot}$. Consequently,
%the variations of $t_a$ are much larger than the variations of
%$t_{\bot}$.
%%GEV
%% Dear Misha, I deleted this because everything depends
%% on what kind of elastic deformation is applied.
%% If the wave-length of deformation exceeds the thickness of the
%% multilayer graphene the displacements of the atoms at any given point $x$

%% the same for all layers (see below).

 First, let us consider the case of the bilayer. We get the Hamiltonian:
\begin{eqnarray}
{\cal H}_2 &=&
\left( \begin{array}{cc}
|{ e}|{\bf e}^{i}_a \sigma^a \circ (\hat{\bf p}_i - {\bf A}_i) &t_{\bot}
\sigma^+\\
t_{\bot}\sigma^-&|{ e}|{\bf e}^{i}_a \sigma^a \circ (\hat{\bf p}_i - {\bf
A}_i) \end{array} \right)
 \nonumber\\ &=&
\left( \begin{array}{cccc}
0  &\{ {\bf E}^{i} \hat{\bf q}_i\}^{\dag}&0&t_{\bot}\\
\{{\bf E}^{i}\hat{\bf q}_i\}&0 &0&0\\
0&0& 0 &\{{\bf E}^{i} \hat{\bf q}_i\}^{\dag}\\
t_{\bot}&0&\{{\bf E}^{i} \hat{\bf q}_i\}&0
\end{array} \right)
\label{2x2}
\end{eqnarray}
We denote
${\bf E}^i = { e} \, ({\bf e}^{i}_1 + i{\bf e}^{i}_2)$ and $\hat{\bf q} = \hat{\bf p} - {\bf A}$. The origin of the
fields $\bf e, A$ is the same as for the case of the monolayer graphene.
In principle, each layer could have its own zweibein $\bf e$, and its own $U(1)$
field $\bf A$.
If the variations of the parameters of the honeycomb lattice are long - wave
(as, for example, in case  of elastic deformations), and the wavelength
exceeds the
thickness of the stack, then connection between the layers can be considered
as rigid, and
the displacements of the atoms at any given point $x$ are
the same for both layers.  As a result the values of the emergent $\bf e$,
$\bf A$ are also the same. In Eq. (\ref{2x2}) the product ${\bf E}^{i} \hat{\bf q}_i$  should be understood as
${\bf E}^{i} \circ \hat{\bf q}_i =\{{\bf E}^{i} \hat{\bf q}_i \}= \frac{1}{2}\Bigl({\bf E}^{i} \overrightarrow{\hat{\bf q}}_i + \overleftarrow{\hat{\bf q}} {\bf E}^i \Bigr)$, where operator $\overrightarrow{\hat{\bf q}}_i$ acts to the right while $\overleftarrow{\hat{\bf q}}_i$ acts to the left. However, the difference between the two products may be neglected as it was explained in Section \ref{sectmono1}.  This allows us to consider the  symbol of the symmetrized product $\circ$ as the usual product if necessary.

One can see, that at ${\bf q} = 0$ there are two degenerate states with zero
energy (correspond to the components $\psi^1_B, \psi^2_A$ ) and two
states with the energies $\pm t_{\bot}$. We call the subspace of low energies $Q$ and the subspace with the high energies $P$.  The high energy hamiltonian is denoted by ${\cal H}_{PP} = t_{\bot} \sigma^1$. The matrix elements of the perturbations between the two subspaces $P$ and $Q$ are given by matrices ${\cal H}_{QP} = {\rm diag} \Bigl(\{{\bf E}^{i} \hat{\bf q}_i\}^{\dag}, \{{\bf E}^{i} \hat{\bf q}_i\} \Bigr)$, and ${\cal H}_{PQ} = {\cal H}_{QP}^{\dag}$.  We consider the terms of the hamiltonian
proportional to ${\bf q}$ as perturbations. The corrections to the energies
of the two degenerate states are given by the second order of the
degenerate state perturbation theory \cite{multilayer}: ${\cal H}_2 \approx - {\cal H}_{QP} {\cal H}_{PP}^{-1} {\cal H}_{PQ}$.  The resulting low energy $2\times 2$ hamiltonian is
given by

\begin{eqnarray}
{\cal H}_2 &\approx&-
\frac{1}{t_{\bot}}\left( \begin{array}{cc}
0  &\Bigl(\{ {\bf E}^{i} \hat{\bf q}_i\}^{\dag}\Bigr)^{2}\\
\{{\bf E}^{i}\hat{\bf q}_i\}^2&0
\end{array} \right)
\label{2x21}
\end{eqnarray}

If the zweibein is constant this would give the dispersion low
$ E^2 = (e^2\sqrt{|g|} g^{ik}q_i q_k)^{2}/t_{\bot}^2$, $g^{ik}={\bf e}^{i}_a
{\bf e}^{k}_b \delta^{ab}$.
For the original choice of the zweibein $|{\bf e}|=1$ on the single brane.
The action would be invariant under the $2D$ - reparametrizations
if we restore $|{\bf e}|$ at the integration measure $d^2x$ and if the
fields $\bf e, A$ are independent.  Then the field - theoretical
Hamiltonian is
\begin{equation}
H = - \frac{1}{t_{\bot}}\int d^2 x  [\psi({\bf x})]^\dag
\Bigl(\sigma^+ \Bigl[e({\bf x})({\bf e}_1+i{\bf e}_2)\circ(\hat{\bf
p}- {\bf A}) \Bigr]^{2}
+
 \sigma^-\Bigl[e({\bf x})({\bf e}_1-i{\bf e}_2)\circ(\hat{\bf p}-{\bf A})\Bigr]^{2}
\Bigr)
\psi({\bf x})\label{D0}
\end{equation}
Function $e(x)$ here is considered as a scalar.

In general case of arbitrary $J>2$ the consideration is similar.
It gives
the following deformation of the gapless branch (for constant ${\bf e}(x)$)
 \begin{equation}
 E^2=(e^2\sqrt{|g|} g^{ik}q_i q_k)^{J}/t_{\bot}^{2(J-1)}~~,~~
g^{ik}={\bf e}^{i}_a{\bf e}^{k}_b \delta^{ab}
 \,,
\label{Hopf_Mapping}
\end{equation}
Recall that $e(x)$ is defined in such a way that ${\rm det} \, {\bf e} = 1$.

The obtained anisotropic
gravity is  based on the original zweibein field ${\bf e}^{i}_a$. The effective theory of
the low-energy gapless fermions is described by the reduced $2\times 2$
Hamiltonian, which
 is expressed in terms of these zweibein
${\bf e}_1$ and ${\bf e}_2$:
\begin{equation}
{\cal H}= (-1)^{J-1}\frac{1}{t_{\bot}^{J-1}} \Bigl(\sigma^+ \Bigl[e({\bf x})({\bf e}_1+i{\bf e}_2)\circ(\hat{\bf
p}- {\bf A}) \Bigr]^{J}
+
 \sigma^-\Bigl[e({\bf x})({\bf e}_1-i{\bf e}_2)\circ(\hat{\bf p}-{\bf A})\Bigr]^{J}
\Bigr)\,.
\label{FermionHamiltonianJ}
\end{equation}
The derivation of Eq. (\ref{FermionHamiltonianJ}) is similar to that of the effective Hamiltonian for the multilayer graphene with ABC stacking presented in Eqs. (2.21) - (2.24) of \cite{multilayer}. It uses the degenerate state perturbation theory. In our case, instead of the operator $v_F\sigma {\bf p}$ we substitute $|{ e}|{\bf e}^{i}_a \sigma^a \circ (\hat{\bf p}_i - {\bf A}_i)$. This does not change the derivation.

The field - theoretical Hamiltonian is
\begin{eqnarray}
H &=& (-1)^{J-1}\frac{1}{t_{\bot}^{J-1}} \int d^2 x |{\rm det}\,{\bf e}^{-1}| [\psi({\bf x})]^\dag
 \Bigl(\sigma^+ \Bigl[e({\bf x})({\bf e}_1+i{\bf e}_2)\circ(\hat{\bf
p}- {\bf A}) \Bigr]^{J}\nonumber\\
&&+
 \sigma^-\Bigl[e({\bf x})({\bf e}_1-i{\bf e}_2)\circ(\hat{\bf p}-{\bf A})\Bigr]^{J}
\Bigr)
\psi({\bf x}),\label{D}
\end{eqnarray}
By construction  $|{\rm det}\,{\bf e}^{-1}| = |{\bf e}|=1$, but we restore in this expression  $|{\bf e}|$ at the integration measure over $\bf x$ in order to demonstrate, that Eq. (\ref{D}) might formally be considered as invariant under the $2D$ reparametrizations if the field $\bf A$ would be independent of $\bf e$ (or absent as in the case when the compensating electromagnetic field is present). For this in $\hat{\bf p}=-i \nabla$ we should substitute the usual derivative via the covariant one. The latter contains the affine connection $\Gamma$ when acts on vectors. This affine connection has zero curvature.  The one - particle hamiltonian
Eq. (\ref{FermionHamiltonianJ}) appears in the (partially) gauge fixed version of Eq.
(\ref{D}) with  $|{\bf e}|=1$, and $\Gamma =0$. In the case, when the fields $\bf e, A$
depend on each other, the introduction of the affine connection is not enough to make the Hamiltonian  invariant under the $2D$ reparametrizations. The independent variables are
the components of the zweibein $\bf e$. The field $\bf A$ is expressed
through $\bf e$. The relation between $\bf A$ and $\bf e$ given by Eq. (\ref{All_inv}) includes tensor $K$ that depends on the orientation of the original honeycomb lattice.  In this situation the theory may be considered
as a gauge fixed version of a rather complicated model with the invariance
under the $2D$ reparametrizations.

The varying $2D$ geometry here is of the Weitzenbock type. The important
property of the resulting theory is its anisotropic scaling with the
parameter of anisotropy $z = J$. Therefore, we call it Horava gravity in
analogy to the theory of quantum gravity with anisotropic scaling considered
in \cite{HoravaPRD2009}. So we came to the non-relativistic extension of the
tetrad gravity  of
Einstein-Cartan-Sciama-Kibble type \cite{Nieh}.

As it was mentioned above, in Eq. (\ref{D}) the product denoted by $\circ$ should be understood as
$\Bigl[e({\bf x})({\bf e}_1+i{\bf e}_2)\circ(\hat{\bf
p}- {\bf A}) \Bigr] = \frac{1}{2}\Bigl(e({\bf x})({\bf e}_1+i{\bf e}_2)(\overrightarrow{\hat{\bf p}} -{\bf A})+ (\overleftarrow{\hat{\bf p}}-{\bf A}) e({\bf x})({\bf e}_1+i{\bf e}_2)\Bigr)$. This product makes the intralayer hamiltonian ${\cal H}_1 = |{ e}|{\bf e}^{i}_a \sigma^a \circ (\hat{\bf p}_i - {\bf A}_i) $ hermitian.  However, according to sect. \ref{sectmono1} within our level of accuracy the difference between this product and the usual one may be neglected, so that in the practical applications of Eq. (\ref{D}) we may also consider this product as the usual one.

\section{Discussion}
\label{DiscussionSection}

First of all, in this paper we reconsider the tight - binding model for the monolayer graphene.   We allow the hopping parameters to vary. In particular, this
model may describe the monolayer graphene with the varying elastic
deformations. We show that  the varying hopping parameters for monolayer graphene give rise to the varying
$2D$ zweibein ${\bf e}^a_k$. The other existing field (the 2D gauge potential
${\bf A}$) is expressed through ${\bf e}$. Therefore, in this case the varying $2D$ Weitzenbock
geometry defined by $\bf e$ appears. The field $\bf A[e]$ gives the
terms of the action Eq. (\ref{Hamiltonian}) that are not invariant under the
$2D$ diffeomorphisms. Formally the
considered action of Eq. (\ref{Hamiltonian}) may
be treated as the action for the $2D$ Weitzenbock
geometry if it is considered as a gauge fixed version
of the action for the invariant theory. This gauge corresponds, in
particular,
to the synchronous reference frame with rescaled time, where $|{\rm det}\,
e|\, e^0_i = \delta^0_i$. Interestingly, if the emergent gauge field is exactly compensated by the external electromagnetic field (for the fermions living to one of the Fermi - points), the effective low energy Hamiltonian obeys the emergent invariance under the $2D$ reparametrizations (near to the given Fermi - point).

In the case of the multilayer graphene with $ABC$ stacking we obtain the similar results. We allow the intralayer hopping parameters to vary but fix the interlayer hopping parameters. The
effective low energy hamiltonian is given by  Eq. (\ref{D}). This is the theory of the varying $2D$
Weitzenbock geometry. Again, there is no emergent invariance under the $2D$
reparametrizations in general case. But the hamiltonian of  Eq.  (\ref{D})
may be considered as the gauge fixed version of the hamiltonian for the
invariant theory.
As for the case of the monolayer graphene, if the emergent gauge field is exactly compensated by the external electromagnetic field (near to the given Fermi - point), the effective low energy Hamiltonian formally obeys the emergent invariance under the $2D$ reparametrizations. (However, for this, unlike the monolayer graphene, we should add the affine connection of zero curvature, and all derivatives are to be understood as the covariant ones.)
The important property of the case $J>1$ is that the
resulting theory obeys anisotropic scaling with $z=J$. Therefore, we
conclude, that the effective low energy theory of multilayer graphene with
the varying intralayer hopping parameters models the Horava - like
teleparallel gravity with anisotropic scaling.

Actually, we do not exclude that  independent contributions to the
effective $SO(2)$ connection and effective $U(1)$ field may appear due to a
mechanism that was not taken into account here.  This would lead to the spin
connection varying independently of the zweibein, and the effective
$U(1)$ field varying independently of the zweibein and of the $SO(2)$
connection. (For example, the dislocations may lead to the independent torsion field.) In this case we would deal with the  emergent Riemann -  Cartan
geometry.

In general, there are two ways to look at the emergent geometry. First of
all, the geometry appears as a result of some deformations of the given
system (for example, of the mechanical elastic deformations). Then, it is
experienced by the propagating fermionic quasiparticles according to Eq.
(\ref{Hamiltonian}), that is the emergent geometry serves as a background
for the fermions. Another look at the emergent geometry is more ambitious.
Suppose, that the fluctuations of the parameters of the given system may
result in the independent fluctuations of the emergent zweibein ${\bf
e}^a_k$. The effective action appears that has, in general, two terms. The
first term $S^{(1)}$ is given by Eq. (\ref{D}) or Eq. (\ref{D0}). The other
term  $S^{(0)}[{\bf e}]$ does not depend on the fermionic fields. Under
certain circumstances $S^{(0)}$ may be neglected, and geometry fluctuates
according to $S^{(1)}$ (after the fermions are integrated out). For this
scenario to work we also need that the integration over fermions is
dominated by the values of momenta for which Eq. (\ref{D}) (or Eq.
(\ref{D0})) is at work. This second scenario is the Sakharov - Zeldovich
scenario of the induced quantum gravity.
In the present paper we imply first of all the first look at the emergent
geometry while the appearance of Sakharov - Zeldovich scenario remains a
subject of a future development. In principle we do not exclude that it can
be realized in monolayer or multilayer graphene,
%%GEV
at least in the form of the subdominant terms in action (see \cite{VolovikZelnikov2003}).

In addition, this pattern may
serve as a tool for the construction of quantum gravity theory in real
space - time. For example, we may consider the $4D$ "graphene" described in
\cite{Creutz}. The variations of the hopping parameters will possibly give
rise to the varying $4D$  geometry. This construction may give an
example of the lattice regularized theory of  quantum gravity. It could be
that the consideration of the stack of such branes with nonzero interlayer
hopping parameter gives another formulation of the lattice regularized
quantum gravity.

The authors kindly acknowledge useful correspondence with M.Vozmediano. 
This work was partly supported by RFBR grant 11-02-01227, by the
Federal Special-Purpose Programme 'Human Capital' of the Russian Ministry of
Science and Education. GEV
acknowledges a financial support of the Academy of Finland and its COE
program,
and the EU  FP7 program ($\#$228464 Microkelvin).

\section{Appendix. Emergent geometry in monolayer graphene}
\label{sectmono}

\subsection{Hamiltonian in momentum space}

Here we demonstrate how the emergent $U(1)$ gauge field and emergent geometry appear as a result of the variations of hopping parameters.
 The carbon atoms of graphene form a
honeycomb
lattice with two sublattices A and B (of the triangular form). Further we
denote the lattice spacing by $a$. Let us introduce vectors that connect a
vertex of the sublattice A to its neighbors (that belong to the sublattice
B):

\begin{equation}
{\bf l}_1= (-a,0),\qquad {\bf l}_2 =  (a/2,a\sqrt{3}/2),\qquad {\bf l}_3=
(a/2,-a\sqrt{3}/2)\label{uuu3}
\end{equation}

Now suppose that the hopping parameter of the tight - binding model varies, so that its value depends on the particular link connecting two adjacent points of the honeycomb lattice. We have three values of $t_a, a = 1,2,3$ at each point.
The Hamiltonian has the form
\begin{equation}
H=-\sum_{\alpha\in A}\sum_{j=1}^3 t_j(\br_\alpha)
    \Bigl(\psi^\dag (\br_\alpha) \psi(\br_\alpha + {\bf l}_j)
        + \psi^\dag (\br_\alpha +{\bf
l}_j)\psi(\br_\alpha)\Bigr)\,,\label{H12}
\end{equation}

We
define the following variables:
\begin{eqnarray}
&&{\bf m}_1= -{\bf l}_1 + {\bf l}_2, \qquad {\bf m}_3 = -{\bf
l}_3 + {\bf l}_1,  \qquad {\bf m}_2=
-{\bf l}_2 + {\bf l}_3 = -{\bf m}_1 - {\bf m}_3
  \label{uuu_2}
\end{eqnarray}
 The effective  Hamiltonian has
has the form
\begin{equation}
H =  \int \frac{d^2k}{\Omega} \frac{d^2k^{\prime}}{\Omega} \psi^\dag({\bf k}^{\prime}) \hat{V}({\bf k}^{\prime},{\bf k}) \psi({\bf
k}),\label{H1_}
\end{equation}
where
\begin{equation}
\hat{V}({\bf k}^{\prime},{\bf k})  = - \sum_{j=1}^3 t_j({\bf k}^{\prime}-{\bf k}) \left(\begin{array}{cc} 0 &   e^{-i{\bf
l}_j {\bf k}} \\
e^{i {\bf l}_j {\bf k}^{\prime}} & 0 \end{array}\right)\label{V1}
\end{equation}
while $\Omega$ is the area of momentum space.

\subsection{Emergent  $U(1)$ field.}

In the further three subsections we consider the situation, when the three hopping parameters $t_a$ are different, but do not depend on the position in coordinate space.
 Therefore,  we substitute ${\bf k} =  {\bf k}^{\prime}$.
The eigenvalues of $\hat{V}$ give the dispersion of the quasiparticles:
\begin{equation}
E(k)=\pm |t_1 + t_2 e^{i {\bf m}_1(x) {\bf k}} + t_3 e^{-i {\bf m}_3(x) {\bf k}} |
\,.\label{EtX}
\end{equation}

We imply, that the variations of $t_a$ are given by
\begin{equation}
t_a({\bf r}) = t (1 - \Delta_a({\bf r})), \quad |\Delta_a| \ll 1
\end{equation}

We  introduce the notation $K^{\pm}$ for the  Fermi point, at which the following expression vanishes:
\begin{equation}
 \sum_{j=1}^3 t_j \left(\begin{array}{cc} 0 &   e^{-i{\bf
l}_j
K^{\pm}} \\
e^{i{\bf l}_j K^{\pm}} & 0 \end{array}\right) = 0
\end{equation}

One can easily find
\begin{equation}
K^{\pm} = \pm \frac{1}{a^2} \Bigl[\frac{\phi_1-\phi_3}{{3}}\,\Bigl(-{{\bf l}_1}\Bigr)+ \frac{\phi_1+\phi_3}{\sqrt{3}}\,\Bigl(-\frac{{\bf m}_2}{\sqrt{3}}\Bigr)\Bigr]\label{FP}
\end{equation}
with
\begin{eqnarray}
\phi_1 & = & \frac{\pi}{2} + {\rm arcsin} \frac{-t_3^2+t_2^2+t_1^2}{2t_2 t_1} \approx \frac{2\pi}{3} + \frac{1}{\sqrt{3}}(2 \Delta_3 - \Delta_1 - \Delta_2) \nonumber\\
\phi_3 & = & \frac{\pi}{2} + {\rm arcsin} \frac{-t_2^2+t_3^2+t_1^2}{2t_3t_1} \approx \frac{2\pi}{3} + \frac{1}{\sqrt{3}}(2 \Delta_2 - \Delta_1 - \Delta_3)
\end{eqnarray}
Recall that vectors $-{\bf l}_1$ and $-\frac{{\bf m}_2}{\sqrt{3}}$ form the orthogonal basis with the length of the basis vectors equal to $a$.
For $t_1=t_2=t_3$ we would have $\phi_1=\phi_3=2\pi/3$, and $K^{(0)}_{\pm} = \mp \frac{4\pi}{9} {\bf m}_2$. With small variations of $t_a$ we get:
\begin{eqnarray}
K^{\pm} &\approx& \pm \frac{1}{a^2}\Bigl[ \frac{\Delta_3-\Delta_2}{\sqrt{3}}\,\Bigl(-{{\bf l}_1}\Bigr)\nonumber\\&& + \Bigl(\frac{4\pi}{3\sqrt{3}}+\frac{\Delta_3+\Delta_2 -
2 \Delta_1}{3}\Bigr)\Bigl(-\frac{{\bf m}_2}{\sqrt{3}}\Bigr)\Bigr]\label{FP2}
\end{eqnarray}

One can see, that in addition to the fixed Fermi - point $K^{(0)}_{\pm} = \pm \frac{4\pi}{3\sqrt{3}}\Bigl(-\frac{{\bf m}_2}{\sqrt{3}}\Bigr)$ the emergent $U(1)$ gauge field ${\bf A}$ appears with the components \cite{sato}:
\begin{eqnarray}
{\bf A}_1  & = &\frac{1}{a}\,\frac{\Delta_3-\Delta_2}{\sqrt{3}}\nonumber\\
  {\bf A}_2  & = & \frac{1}{a}\,\frac{\Delta_3+\Delta_2 -
2 \Delta_1}{3}\label{AFP2}
\end{eqnarray}

\subsection{Emergent  zweibein: expansion around the true Fermi - point}
We are going to expand $\hat{V}$ around the true Fermi - point. So, we set $\bk  = K_\pm +\bq$. The result will be presented in the form:
\begin{eqnarray}
 \hat{V}_\pm&=& (\pm \sigma^1 {\bf f}_2 + \sigma^2 {\bf f}_1  ){\bf q},\label{Hpm2}
\end{eqnarray} 
where $\bf f$ is to be defined below. 
Let us denote
\begin{equation}
\hat{V}  = \left(\begin{array}{cc} 0 &  U  \\
U^+  & 0 \end{array}\right), \quad U = - t \sum_{j=1}^3 (1 - \Delta_j) e^{-i{\bf
l}_j {\bf k}} \label{V2_}
\end{equation}

Next, we expand $U$
around $K^{\pm}$:
\begin{eqnarray}
U  &=&  - t \sum_{j=1}^3 (1 - \Delta_j) e^{-i{\bf
l}_j (K_\pm +\bq)}\nonumber\\&\approx & - t \sum_{j=1}^3 (1 - \Delta_j) e^{-i{\bf
l}_j K_\pm }(1 - i\, {\bf
l}_j \bq) \nonumber\\
&=&    i\, t \sum_{j=1}^3 (1 - \Delta_j) e^{-i{\bf
l}_j K_\pm }\, {\bf
l}_j \bq = {\bf F}_{\pm} \, \bq,\label{V2__}
\end{eqnarray}
where 
\begin{equation}
{\bf F}_{\pm} =    i\, t \sum_{j=1}^3 (1 - \Delta_j) e^{-i{\bf
l}_j K_\pm }\, {\bf
l}_j 
\end{equation}

One can see, that this tensor is related to the coefficients $\bf f$ in the expansion of Eq. (\ref{Hpm2}) as follows:
\begin{equation}
{\bf F}_{\pm} = \pm {\bf f}_2 - i {\bf f}_1,
\end{equation}
so that ${\bf F}_{-} = -{\bf F}^*_{+}$.

Next, recall,  that $K_{\pm} = K^{(0)}_{\pm} \pm {\bf A}$, and $K^{(0)}_{\pm} = \pm \frac{4\pi}{3\sqrt{3}}\Bigl(-\frac{{\bf m}_2}{\sqrt{3}}\Bigr)= \pm K^{(0)}$. Therefore,

\begin{eqnarray}
{\bf F}_{+}& = &   i\, t \sum_{j=1}^3 (1 - \Delta_j)(1 - i\, {\bf
l}_j  {\bf A} ) e^{-i{\bf
l}_j K^{(0)} }\, {\bf
l}_j \nonumber\\ & \approx &   i\, t \sum_{j=1}^3 (1 - \Delta_j - i\, {\bf
l}_j  {\bf A} ) e^{-i{\bf
l}_j K^{(0)} }\, {\bf
l}_j \label{FA}
\end{eqnarray}

As a result
\begin{eqnarray}
{\bf f}_{2}& = &    t \sum_{j=1}^3 \Bigl((1 - \Delta_j) {\rm sin} ({\bf
l}_j K^{(0)} ) + {\bf
l}_j  {\bf A} \, {\rm cos} ({\bf
l}_j K^{(0)} ) \Bigr) \, {\bf
l}_j \nonumber\\
{\bf f}_{1}& = &    t \sum_{j=1}^3 \Bigl(-(1 - \Delta_j) {\rm cos} ({\bf
l}_j K^{(0)} ) + {\bf
l}_j  {\bf A} \, {\rm sin} ({\bf
l}_j K^{(0)} ) \Bigr) \, {\bf
l}_j
\end{eqnarray}

We use, that ${\bf l}_1 K^{(0)} = 0$, ${\bf l}_2 K^{(0)} =  \frac{2\pi}{3}$, ${\bf l}_3 K^{(0)} =  -\frac{2\pi}{3}$, and get:
\begin{eqnarray}
{\bf f}_{2}& = &    t \sum_{N=0, \pm 1} \Bigl((1 - \Delta_{j[N]}) \frac{\sqrt{3}}{2} N  + {\bf
l}_{j[N]}  {\bf A} \, (1-3 |N|/2) \Bigr) \, {\bf
l}_{j[N]} \nonumber\\
 {\bf f}_{1}& = &    t \sum_{N = 0, \pm 1}^3 \Bigl(-(1 - \Delta_{j[N]})(1-3 |N|/2) + {\bf
l}_{j[N]}  {\bf A} \, \frac{\sqrt{3}}{2} N) \Bigr) \, {\bf
l}_{j[N]}
\end{eqnarray}
where $j[N] = 1 -N/2 + 3 |N|/2$. Next, we rewrite this expression in tensorial form:
\begin{eqnarray}
{\bf f}^k_{a}& = &    \frac{t}{a} \sum_{j} \Bigl((1 - \Delta_{j}) {\bf l}_j^a   + ({\bf
l}_{j}  {\bf A}) \,  \epsilon^{ab} {\bf l}_j^b \Bigr) \, {\bf
l}_{j}^k \nonumber\\
& = &    \frac{t}{a} \sum_{j} {\bf l}_j^a  {\bf l}_j^k -\frac{t}{a}\sum_j  \Delta_{j} {\bf l}_j^a {\bf l}_j^k    + \frac{t}{a}\epsilon^{ab} {\bf A}^m \, \sum_j {\bf
l}^m_{j}  {\bf l}_j^b {\bf
l}_{j}^k
\end{eqnarray}

Let us introduce the new tensor $K$:
\begin{eqnarray}
K^{ijk}=-\frac{4}{3a^3} \sum_b {\bf l}_b^i {\bf l}^j_b {\bf l}_b^k, \quad  K^{111}=-K^{122}=-K^{221}=-K^{212} = 1
\end{eqnarray}
(Tensor $K$ was first introduced in \cite{Vozmediano}. It reflects the structure of the honeycomb lattice.)
Also we use the relation
\begin{eqnarray}
\sum_b {\bf l}_b^i {\bf l}^j_b  &=& a^2 \frac{3}{2} \delta^{ij}
\end{eqnarray}

Expression for $\bf f$ receives the form:
\begin{eqnarray}
{\bf f}^k_{a}& = &    \frac{3ta}{2}\Bigl( \delta^k_a  - \frac{2}{3a^2} \sum_j  \Delta_{j} {\bf l}_j^a {\bf l}_j^k    - \frac{a}{2}\epsilon^{ab} {\bf A}^m \, K^{mbk} \Bigr)
\end{eqnarray}

The equation for the field $\bf A$ may be represented in the same spirit:
\begin{eqnarray}
 0&=&  -\frac{t}{a}\sum_j  \Delta_{j} {\bf l}_j^a     + \frac{t}{a}\epsilon^{ab} {\bf A}^m \, \sum_j {\bf
l}^m_{j}  {\bf l}_j^b \nonumber\\
0&=&  -\sum_j  \Delta_{j} {\bf l}_j^a     + \frac{3 a^2}{2}\epsilon^{ab} {\bf A}^b
\nonumber\\
{\bf A}^b &=&  - \frac{2}{3a^2} \epsilon^{ba} \sum_j  \Delta_{j} {\bf l}_j^a
\end{eqnarray}

\subsection{Emergent  geometry plus emergent $U(1)$ field}

We substitute the expression for $\bf A$ into the expression for $\bf f$, and obtain:
\begin{eqnarray}
{\bf f}^k_{a}& = &    v_F \Bigl( \delta^k_a  - \frac{2}{3a^2} \sum_j  \Delta_{j} \Bigl[{\bf l}_j^a {\bf l}_j^k    - \frac{a}{2} {\bf l}^d_j \, K^{dak}\Bigr] \Bigr)
\end{eqnarray}

This gives
\begin{eqnarray}
 {\bf f}_1
&\approx& t \sqrt{3}\frac{(-\Delta_2 +  \Delta_3)}{2}\Bigl(-\frac{{\bf m}_2}{\sqrt{3}}\Bigr)+ \frac{3t}{2}(1-\Delta_1)(-{\bf l}_1) \nonumber\\
 {\bf f}_2 &\approx&  \frac{3t}{2}(1 - \frac{1}{3}(2\Delta_2+2\Delta_3-\Delta_1))\Bigl(-\frac{{\bf m}_2}{\sqrt{3}}\Bigr) + t\sqrt{3} \frac{(-\Delta_2 +  \Delta_3)}{2} (-{\bf l}_1) \label{egraph12}
\end{eqnarray}

One can see, that the mentioned above $U(1)$ field is related to the field $\bf f$ as follows:
\begin{equation}
{\bf A}_1  =  \frac{1}{2v_F a} ({\bf f}^1_2 + {\bf f}^2_1), \quad {\bf A}_2 = \frac{1}{2v_F a} ({\bf f}^1_1 - {\bf f}^2_2),\label{a}
\end{equation}
that is 
\begin{eqnarray}
{\bf A}^i & = &  -\frac{1}{2v_Fa}\epsilon^{ik} K^{kjb} {\bf f}^j_b
\end{eqnarray}

The field $\bf f$ has the form:

\begin{equation}
{\bf f}^i_a = v_F\left(\delta^i_a -  \left[\begin{array}{cc} \Delta_{1} &  \frac{(\Delta_2 -  \Delta_3)}{\sqrt{3}} \\
                                                           \frac{(\Delta_2 -  \Delta_3)}{\sqrt{3}} &   \frac{1}{3}(2\Delta_2+2\Delta_3-\Delta_1)   \end{array}\right]\right)\label{f20}
\end{equation}

Close to the Fermi - points we define the new spinors:
\begin{equation}
\Psi_{\pm}({\bf Q}) = \psi(K^{(0)}_{\pm} + {\bf Q})
\end{equation}

As a result, the effective  Hamiltonian has the form:
\begin{equation}
H =  \int \frac{d^2{\bf Q}}{\Omega}  \Psi^\dag({\bf Q}) \hat{V}_{\pm}({\bf Q}) \Psi({\bf
Q}),\label{H1_Q}
\end{equation}
where
\begin{eqnarray}
 \hat{V}_\pm &=&- i \sigma^3\Bigl[(\mp \sigma^2 {\bf f}_2 + \sigma^1 {\bf f}_1  )\Bigl({\bf Q}\mp{\bf A}\Bigr)\Bigr],\label{Hpm2Q}
\end{eqnarray}

Next, we return to the coordinate space using the rule
\begin{eqnarray}
{\bf Q}&\rightarrow & - i  {\nabla}
\end{eqnarray}

 As a result the hamiltonian has the form
\begin{equation}
H =  \sum_{\pm}\int d^2 x [\Psi^{\pm}({\bf x})]^\dag {\bf H}_{\pm}
\Psi^{\pm}({\bf x}),\label{Hamiltonian2}
\end{equation}
where
\begin{eqnarray}
{\bf H}_- &=&  - \sigma^3\, {\bf f}_a^k \sigma^a
[\partial_k + i {\bf A}_k],\quad  a=1,2; k = 1,2;\nonumber \\
{\bf H}_+  &=&- \sigma^2 \Bigl( \sigma^3 \, {\bf f}_a^k \sigma^a
[\partial_k - i {\bf A}_k ]\Bigr) \sigma^2.\label{Hpm3}
\end{eqnarray}
Here the field $\bf f$ is defined in coordinate space and is related to the variables $\Delta_a$ according to  Eq. (\ref{f20}). The field $\bf A$ is given by Eq. (\ref{a}).

The field ${\bf f}$ is related to the dreibein ${\bf e}$ as follows:
\begin{equation}
e \,{\bf e}_a^i = {\bf f}_a^i; \quad e\, {\bf e}_0^0 = 1, \quad {\bf e}_a^0 = {\bf e}_0^i = 0,  \, {\rm where} \,  i,a=1,2
\end{equation}
Here the determinant of the zweibein ${\rm det}\, {\bf e}_a^k = 1$, where $a,k = 1,2$. At the same time the three - dimensional determinant of ${\bf e}$ is equal to ${\rm det}\, {\bf e}^{(3\times3)} = {\bf e}^0_0 = 1/e$.  The three - volume element is $d^{(3)} V = d^2 {\bf r} d t e({\bf r},t)$. The value of $e$ is given by
\begin{equation}
e = [{\rm det}\, {\bf f}]^{1/2} = v_F(1- \frac{1}{3}(\Delta_2 + \Delta_3 + \Delta_1))
\end{equation}

\subsection{Emergent  geometry plus emergent $U(1)$ field minus real electromagnetic field}

Let us suppose, that the emergent gauge field Eq. (\ref{AFP2}) is exactly compensated by the external electromagnetic field for the fermions living near to the Fermi - point $K_-$.
In this situation instead of Eq. (\ref{V1}) we should use the following expression:
\begin{equation}
\hat{V}({\bf k})  = - \sum_{j=1}^3 t_j({\bf k} - {\bf k}^{\prime}) \left(\begin{array}{cc} 0 &   e^{-i{\bf
l}_j
({\bf k}+{\bf A})} \\
e^{i{\bf l}_j ({\bf k}^{\prime}+{\bf A})} & 0 \end{array}\right)\label{V1_},
\end{equation}
 where $\bf A$ is the external electromagnetic field that compensates exactly the emergent $U(1)$ field for $\Psi_-$. (Further in this subsection ${\bf k} = {\bf k}^{\prime}$.)
We set
$\bk  = K^{(0)}_\pm +\bq$,  and expand $\hat{V}_-({\bf k})$
around $K_{-}^{(0)}$. In practise this means that we expand $V_-({\bf k})$ around the Fermi point because in this case $V_-({\bf K}_{-}^{(0)}) = 0$:
\begin{eqnarray}
 \hat{V}_-&=& (- \sigma^1 {\bf f}_2 + \sigma^2 {\bf f}_1  ){\bf k},\label{Hpm22}
\end{eqnarray}

Again, close to the Fermi - points we define the new spinors:
\begin{equation}
\Psi_{\pm}({\bf Q}) = \psi(K^{(0)}_{\pm} + {\bf Q})
\end{equation}

In coordinate space the hamiltonian has the form
\begin{equation}
H =  \sum_{\pm}\int d^2 x [\Psi^{\pm}({\bf x})]^\dag {\bf H}_{\pm}
\Psi^{\pm}({\bf x}),\label{Hamiltonian2_}
\end{equation}
where
\begin{eqnarray}
{\bf H}_- &=& - \sigma^3\, {\bf f}_a^k \sigma^a
\partial_k\nonumber \\
{\bf H}_+  &=& - \sigma^2 \Bigl( \sigma^3 \, {\bf f}_a^k \sigma^a[\partial_k - 2 i {\bf A}_k ]
\Bigr) \sigma^2\label{Hpm6}
\end{eqnarray}
Here the field ${\bf f}$ is defined in coordinate space and is related to the variables $\Delta_a$ according to  Eq. (\ref{f20}).

\subsection{Inhomogenious hopping parameters}
\label{Appendix5}

In the case of the hopping parameters depending on the position in coordinate space we need to use expression Eqs. (\ref{V1})
and (\ref{V1_}) with ${\bf k} \ne {\bf k}^{\prime}$.

As a result, the effective  Hamiltonian has the form:
\begin{equation}
H =  \int \frac{d^2{\bf Q}}{\Omega} \frac{d^2{\bf Q}^{\prime}}{\Omega} \Psi^\dag({\bf Q}^{\prime}) \hat{V}_{\pm}({\bf Q}^{\prime},{\bf Q}) \Psi({\bf
Q}),\label{H1_Q}
\end{equation}
where
\begin{eqnarray}
 \hat{V}_\pm({\bf Q},{\bf Q}^{\prime})&=& -i \sigma^3\Bigl[(\mp \sigma^2 {\bf f}_2 + \sigma^1 {\bf f}_1  )\Bigl(\frac{{\bf Q}+{\bf Q}^{\prime}}{2}\mp{\bf A}\Bigr)
  - ( \sigma^1 {\bf f}_1 \mp \sigma^2 {\bf f}_2  ) \sigma^3  \frac{{\bf Q}-{\bf Q}^{\prime}}{2}\Bigr],\label{Hpm2Q}
\end{eqnarray}
(Below for the definiteness we consider the case, when there is no compensating electromagnetic field.)
In this case in order to return to the coordinate space we should use the following rule
\begin{eqnarray}
\frac{1}{2} ({\bf Q} + {\bf Q}^{\prime}) F({\bf Q} - {\bf Q}^{\prime})&\rightarrow & -\frac{i}{2} ( F({\bf x}) \overrightarrow{\nabla} - \overleftarrow{\nabla} F({\bf x}))\nonumber\\
\frac{1}{2} ({\bf Q} - {\bf Q}^{\prime}) F({\bf Q} - {\bf Q}^{\prime})&\rightarrow & - \frac{i}{2} \Bigl(\nabla F({\bf x})\Bigr)
\end{eqnarray}

As in the previous subsections, the hamiltonian has the form
\begin{equation}
H =  \sum_{\pm}\int d^2 x [\Psi^{\pm}({\bf x})]^\dag {\bf H}_{\pm}
\Psi^{\pm}({\bf x}),\label{Hamiltonian2__}
\end{equation}
But now
\begin{eqnarray}
{\bf H}_- &=&  -\sigma^3\, {\bf f}_a^k({\bf x}) \sigma^a\circ
[\partial_k + i ({\bf A}_k({\bf x}) +  \tilde{\bf A}_k({\bf x}))]\nonumber \\
{\bf H}_+  &=&- \sigma^2 \Bigl( \sigma^3 \, {\bf f}_a^k({\bf x}) \sigma^a\circ
[\partial_k - i ({\bf A}_k({\bf x}) +  \tilde{\bf A}_k)]\Bigr) \sigma^2\label{Hpm33}
\end{eqnarray}
Here the field $\bf f$ is defined in coordinate space and is related to the variables $\Delta_a$ according to  Eq. (\ref{AFP2}). The additional gauge field $\tilde{\bf A}$ appears:
\begin{eqnarray}
\tilde{\bf A}_a({\bf x}) &=&  \frac{1}{2v_F} \nabla_i{\bf f}^i_b({\bf x}) \epsilon_{ba}\label{c}
\end{eqnarray}

The additional field $\tilde{\bf A}$ is to be compared with the emergent gauge field $\bf A$.
One can see that $\tilde{\bf A} \sim a \nabla {\bf A}$. Therefore, this is not reasonable to keep this additional field together with $\bf A$ in the field - theoretical description, where all dimensional quantities are to be much larger than the lattice spacing $a$. This shows that even in case of the variations of $t_a$ depending on the position in coordinate space we may omit the derivatives of $t_a(x)$ in the effective low energy field - theoretical Hamiltonian. This means that we may formally use the expressions of Eqs. (\ref{Hpm3}) and (\ref{Hpm6}) substituting there the fields $\bf f$, $\bf A$ depending on the coordinates.  The product $i {\bf f}^k_a \circ \partial_k$ in these equations should be understood as
\begin{equation}
{\bf f}^k_a \circ i \partial_k = \frac{i}{2} \Bigl( {\bf f}^k_a  \overrightarrow{\partial_k} - \overleftarrow{\partial_k} {\bf f}^k_a \Bigr)
\end{equation}

However, the difference  ${\bf f}^k_a \circ i \partial_k - {\bf f}^k_a  \partial_k = \frac{i}{2} \Bigl( {\partial_k} {\bf f}^k_a \Bigr) $ may be neglected as well for the same reason as  the field $\tilde{\bf A}$.

\end{document}